# Predicting Group Evolution in the Social Network

Piotr Bródka, Przemysław Kazienko, Bartosz Kołoszczyk

Institute of Informatics, Wrocław University of Technology, Wrocław, Poland
`piotr.brodka@pwr.wroc.pl, kazienko@pwr.wroc.pl,`
`bkoloszczyk@gmail.com`

**Abstract.** Groups – social communities are important components of entire societies, analysed by means of the social network concept. Their immanent feature is continuous evolution over time. If we know how groups in the social network has evolved we can use this information and try to predict the next step in the given group evolution. In the paper, a new aproach for group evolution prediction is presented and examined. Experimental studies on four evolving social networks revealed that (i) the prediction based on the simple input features may be very accurate, (ii) some classifiers are more precise than the others and (iii) parameters of the group evolution extracion method significantly influence the prediction quality.

**Keywords :** social network, group evolution, predicting group evolution, group dynamics, social network analysis, *GED*

## 1   Introduction and Related Work

In most fields of science, researchers struggle to predict the future: the future consumption of power in electric network, future load of network grid, future consumption of goods etc. Social networks are no different. Recently, the main focus is on the link prediction [13], but there are also papers on (i) entire network structure modelling [18], (ii) modelling social network evolution [12], [15], or (iii) churn prediction and its influence on the network [10], [19]. However only few researchers have considered groups in the prediction process. Some of them like Zheleva et. al. are using communities only for link prediction [20], the others like Kairam et. al. tries to identify and understand the factors contributing in the growth and longevity of groups within social networks [9]. Unfortunately, there is no research directly regarding prediction of the entire group evolution. Probably, the main reason behind this is the fact that the methods for determining group history have not been good enough so far.

The approach presented in this paper, involves usage of the results produced by the *GED* method [3] to predict group evolution. The assumption is that using the information about preceding changes of a given group and its characteristic in the past, as the input for the classifier, which was previously trained based on the historical changes of other groups in the social network, we can try to predict the next step in the given group evolution. Based on this assumption, a new approach for group evolution prediction was develop and it is presented and examined in this paper. The results of the first experiments on four evolving social networks revealed that (1) the prediction based on the proposed input features may be very accurate, (2) some classifiers like C4.5 decision trees or random forests are more precise than the others and (3) parameters of the group evolution identification method (GED) [3] significantly influence on the prediction quality.

## 2  *GED* Group Evolution Discovery

The concept of *GED* method and its full evaluation was presented in [3]. In this paper only the most important elements are presented, in order to help the reader understand the next chapters.

### 2.1  Temporal Social Network and Groups

Temporal social network *TSN* is a list of following timeframes (time windows) *T*. Each timeframe is in fact social network *SN(V,E)* where: *V* – is a set of vertices and *E* is a set of directed edges $<x,y>:x,y \in V$

$$\begin{aligned}
TSN &= <T_1, T_2, \ldots, T_m>, \quad m \in N \\
T_i &= SN_i(V_i, E_i), \quad i = 1, 2, \ldots, m \\
E_i &= <x, y>: x, y \in V_i, \quad i = 1, 2, \ldots, m
\end{aligned} \quad (1)$$

### 2.2  Group Evolution

Group evolution is a sequence of events (changes) succeeding each other in the consecutive time windows (timeframes) within the social network. Possible events in social group evolution are:

1. *Continuing* (stagnation), when groups in the consecutive time windows are identical or when groups differ only by few nodes and their size remains the same.
2. *Shrinking*, when nodes has left the group, making its size smaller than in the previous time window. Like in case of growing, a group can shrink slightly as well as greatly.
3. *Growing* (opposite to shrinking), when new nodes has joined to the group, making its size bigger than in the previous time window. A group can grow slightly as well as significantly, doubling or even tripling its size.

4. *Splitting* occurs, when a group splits into two or more groups in the next time window. Like in merging, we can distinguish two types of splitting: equal and unequal, which might be similar to shrinking.
5. *Merging*, (reverse to splitting) when a group consist of two or more groups from the previous time window. Merge might be (1) *equal*, which means the contribution of the groups in merged group is almost the same, or (2) unequal, when one of the groups has much greater contribution into the merged group. In second case merging might be similar to growing.
6. *Dissolving*, when a group ends its life and does not occur in the next time window.
7. *Forming* of new group, which has not exist in the previous time window. In some cases, a group can be inactive over several timeframes, such case is treated as dissolving of the first group and forming again of the second one.

### 2.3 GED – a Method for Group Evolution Discovery in the Social Network

To discover group evolution in the social network a method called *GED* (Group Evolution Discovery) was used [3]. The most important component of this method is a measure called inclusion. This measure allows to evaluate the inclusion of one group in another. Therefore, inclusion $I(G_1,G_2)$ of group $G_1$ in group $G_2$ is calculated as follows:

$$I(G_1, G_2) = \overbrace{\frac{|G_1 \cap G_2|}{|G_1|}}^{\text{group quantity}} \cdot \underbrace{\frac{\sum_{x \in (G_1 \cap G_2)} NI_{G_1}(x)}{\sum_{x \in (G_1)} NI_{G_1}(x)}}_{\text{group quality}} \qquad (2)$$

where $NI_{G_1}(x)$ is the value reflecting importance of the node $x$ in group $G_1$.

As a node importance $NI_{G_1}(x)$ measure, any metric which indicate member position within the community can be used, e.g. centrality degree, betweenness degree, page rank, social position etc. The second factor in Equation 2 would have to be adapted accordingly to selected measure.

The *GED* method, used to discover group evolution, respects both the quantity and quality of the group members. The *quantity* is reflected by the first part of the *inclusion* measure, i.e. what portion of members from group $G_1$ is in group $G_2$, whereas the *quality* is expressed by the second part of the *inclusion* measure, namely what contribution of important members from group $G_1$ is in $G_2$. It provides a balance between the groups that contain many of the less important members and groups with only few but key members. The procedure for the *GED* is as follows:

> **Input:** Temporal social network *TSN*, in which groups are extracted by any community detection algorithm separately for each timeframe $T_i$ and any node importance measure is calculated for each group.

1. For each pair of groups <$G_1$, $G_2$> in consecutive timeframes $T_i$ and $T_{i+1}$ inclusion $I(G_1,G_2)$ for $G_1$ in $G_2$ and $I(G_2,G_1)$ for $G_2$ in $G_1$ is computed

2. Based on both inclusions $I(G_1,G_2)$, $I(G_2,G_1)$ and sizes of both groups only one type of event may be identified:

   a. *Continuing*: $I(G_1,G_2) \geq \alpha$ and $I(G_2,G_1) \geq \beta$ and $|G_1| = |G_2|$

   b. *Shrinking*: $I(G_1,G_2) \geq \alpha$ and $I(G_2,G_1) \geq \beta$ and $|G_1| > |G_2|$ OR $I(G_1,G_2) < \alpha$ and $I(G_2,G_1) \geq \beta$ and $|G_1| \geq |G_2|$ OR $I(G_1,G_2) \geq \alpha$ and $I(G_2,G_1) < \beta$ and $|G_1| \geq |G_2|$ and there is only one match between $G_1$ and groups in the next time window $T_{i+1}$

   c. *Growing*: $I(G_1,G_2) \geq \alpha$ and $I(G_2,G_1) \geq \beta$ and $|G_1|<|G_2|$ OR $I(G_1,G_2) \geq \alpha$ and $I(G_2,G_1) < \beta$ and $|G_1| \leq |G_2|$ OR $I(G_1,G_2) < \alpha$ and $I(G_2,G_1) \geq \beta$ and $|G_1| \leq |G_2|$ and there is only one match between $G_2$ and groups in the previous time window $T_i$

   d. *Splitting*: $I(G_1,G_2) < \alpha$ and $I(G_2,G_1) \geq \beta$ and $|G_1| \geq |G_2|$ OR $I(G_1,G_2) \geq \alpha$ and $I(G_2,G_1) < \beta$ and $|G_1| \geq |G_2|$ and there is more than one match between $G_1$ and groups in the next time window $T_{i+1}$

   e. *Merging*: $I(G_1,G_2) \geq \alpha$ and $I(G_2,G_1) < \beta$ and $|G_1| \leq |G_2|$ OR $I(G_1,G_2) < \alpha$ and $I(G_2,G_1) \geq \beta$ and $|G_1| \leq |G_2|$ and there is more than one match between $G_2$ and groups in the previous time window $T_i$

   f. *Dissolving*: for $G_1$ in $T_i$ and each group $G_2$ in $T_{i+1}$ $I(G_1,G_2) < 10\%$ and $I(G_2,G_1) < 10\%$

   g. *Forming*: for $G_2$ in $T_{i+1}$ and each group $G_1$ in $T_i$ $I(G_1,G_2) < 10\%$ and $I(G_2,G_1) < 10\%$

For more detailed description of *GED* Method and its evaluation see [3].

## 3  The Concept of Using the *GED* Method for Prediction of Group Evolution

Presented approach, involves usage of the results of *GED* method. The assumption is that using a simple sequence, which consists only of several preceding groups' sizes and events, as an input for the classifier, the learnt model will be able to produce very good results even for simple classifiers.

The sequences of groups sizes and events between timeframes can be extracted from the *GED* results. In this paper 4-step sequences were used (Figure 1). Obviously, the event types varied depending on the individual groups, but the time frame numbers were fixed. It means that for each event four group profiles in four previous time frames together with three associated events were identified as the input for the classification model, separately for each group. A single group in a given time frame ($T_n$) was a case (instance) for classification, for which its event $T_nT_{n+1}$ was predicted.

The sequence presented in Figure 1 was used as an input for classification. The first part of the sequence was used as 7 input features (variables), i.e. (1) **Group size** in $T_{n-3}$, (2) **Event type $T_{n-3}T_{n-2}$**, (3) **Group size in $T_{n-2}$**, (4) **Event type $T_{n-2}T_{n-1}$**, (5) **Group size in $T_{n-1}$**, (6) **Event type $T_{n-1}T_n$**, (7) **Group size in $T_n$**. A predictive variable was the next event for a given group. Thus, the goal of classification was to predict (classify) **Event $T_nT_{n+1}$ type** – out of the six possible classes: i.e. (1) growing, (2) continuing, (3) shrinking, (4) dissolving, (5) merging and (6) splitting. Forming was excluded since it can only start the sequence.

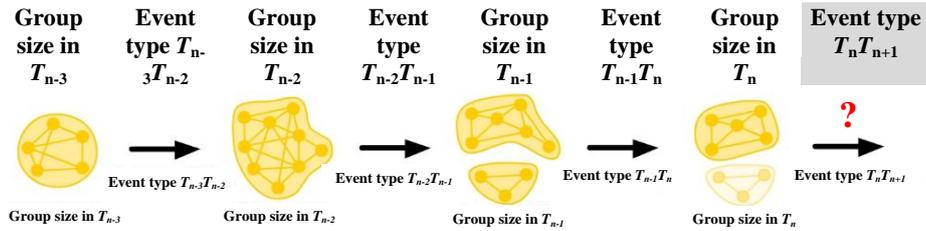

**Fig. 1.** The sequence of events for a single group together with its intermediate sizes (descriptive input variables) as well as its target class - event type in $T_nT_{n+1}$. It corresponds to one case in classification

## 4   Experiment Setup

As mentioned before, the notion, which was checked during the experiments, was that using the information about preceding changes of a given group as well as its description in the past as the input for the classifier, trained based on the historical transitions of other groups, we can try to predict the next step in the given group evolution.

To check this four temporal social networks *TSN* have been extracted from four different datasets to perform and evaluate prediction of group evolution.

1. The first network was extracted from Wroclaw University of Technology email communication. The whole data set was collected within the period from February 2006 to October 2007 and consists of 5,845 nodes (distinct university employees' email addresses) and 149,344 edges (emails send from one address to another). The temporal social network consisted of fourteen 90-days timeframes extracted from this source data. Timeframes have the 45-days overlap, i.e., the first timeframe begins on the 1st day and ends on the 90th day, the second begins on the 46th day and ends on the 135th day and so on.
2. The third social network was extracted from the portal www.salon24.pl, which is dedicated especially to political discussions, but also some other subjects from different domains may be brought up there. The network consists of 3,775 nodes and 77,932 edges. There are 12 non-overlapping timeframes representing 12 months of the 2009 year.
3. The forth one is the well-known Enron e-mail network with 150 nodes and 2,144 edges. The network was split into twelve, 90-days timeframes without overlap.

4. The fifth network was extracted from the portal extradom.pl. It gathers people, who are engaged in building their own houses in Poland. It helps them to exchange best practices, experiences, evaluate various constructing projects and technologies or simply to find the answers to their questions provided by others. The data covers a period of 17 months and contains 3,690 users and 34,082 relations. 33 timeframes were extracted, each of them 30 days long with 15 days overlap, similarly to the first data set.

For each timeframe social communities were extracted using CFinder [16] and for each *TSN* the *GED* method [3] was utilized to extract groups evolution. The *GED* method was run 36 times for each *TSN* with all combination of α and β parameters from the set {50%, 60%, 70%, 80%, 90%, 100%}. As a node importance measure the social position measure [21] (measure similar to page rank) was utilized.

Next, the 4-step sequences where separately extracted from the *GED* results for all networks and every combination of α and β parameters, see an example sequence in Figure 1.

Experiment was performed in WEKA Data Mining Software [7]. Ten different classifiers were utilized with default settings: (see Table 1). For the method of validation 10-fold cross-validation was utilized as the most commonly used [14]. In WEKA, this means 100 calls of one classifier with training data and tested against the test data in order to get statistically meaningful results.

**Table 1.** WEKA classifiers used

| WEKA name | Name |
|---|---|
| BayesNet | Bayes Network classifier [7] |
| NaiveBayes | Naive Bayesian classifier [8] |
| IBk | k-nearest neighbor classifier [1] |
| KStar | Instance-Based classifier [4] |
| AdaBoost | Adaboost M1 method [6] |
| DecisionTable | Decision table [11] |
| JRip | RIPPER rule classifier [5] |
| ZeroR | 0-R classifier |
| J48 | C4.5 decision tree [17] |
| RandomForest | Random forest [2] |

## 5  Results

All classifiers were utilized for each of 4 networks and each combination of α and β parameters. The measure selected for presentation and analysis of the results is F measure which is the harmonic mean of precision and recall.

At the beginning, the classifiers were compared for each dataset separately in order to indicate which one is the best. The results are presented in Table 2 and Figures 2-5.

**Table 2.** The classifiers comparison for each dataset

| Data set | Classifier | Max F measure | Min F measure | Diff. |
|---|---|---|---|---|
| salon24.pl | BayesNet | 1.00 | 1.00 | 0.00 |
| | NaiveBayes | 1.00 | 1.00 | 0.00 |
| | IBk | 1.00 | 1.00 | 0.00 |
| | KStar | 1.00 | 1.00 | 0.00 |
| | AdaBoostM1 | 1.00 | 0.70 | 0.30 |
| | DecisionTable | 1.00 | 0.90 | 0.11 |
| | JRip | 1.00 | 0.97 | 0.03 |
| | ZeroR | 0.82 | 0.60 | 0.23 |
| | J48 | 1.00 | 0.99 | 0.01 |
| | RandomForest | 1.00 | 1.00 | 0.00 |
| Enron | BayesNet | 0.83 | 0.69 | 0.15 |
| | NaiveBayes | 0.81 | 0.72 | 0.08 |
| | IBk | 0.79 | 0.71 | 0.08 |
| | KStar | 0.79 | 0.72 | 0.07 |
| | AdaBoostM1 | 0.51 | 0.32 | 0.20 |
| | DecisionTable | 0.78 | 0.64 | 0.14 |
| | JRip | 0.80 | 0.73 | 0.07 |
| | ZeroR | 0.27 | 0.15 | 0.11 |
| | J48 | 0.92 | 0.80 | 0.13 |
| | RandomForest | 0.89 | 0.76 | 0.13 |
| extradom.pl | BayesNet | 0.87 | 0.54 | 0.32 |
| | NaiveBayes | 0.87 | 0.50 | 0.37 |
| | IBk | 0.88 | 0.55 | 0.33 |
| | KStar | 0.88 | 0.52 | 0.36 |
| | AdaBoostM1 | 0.83 | 0.50 | 0.33 |
| | DecisionTable | 0.88 | 0.48 | 0.39 |
| | JRip | 0.88 | 0.35 | 0.53 |
| | ZeroR | 0.88 | 0.33 | 0.54 |
| | J48 | 0.88 | 0.33 | 0.55 |
| | RandomForest | 0.88 | 0.40 | 0.48 |
| WrUT emails | BayesNet | 0.86 | 0.76 | 0.10 |
| | NaiveBayes | 0.86 | 0.73 | 0.13 |
| | IBk | 0.88 | 0.79 | 0.09 |
| | KStar | 0.88 | 0.81 | 0.08 |
| | AdaBoostM1 | 0.68 | 0.54 | 0.14 |
| | DecisionTable | 0.88 | 0.74 | 0.14 |
| | JRip | 0.83 | 0.78 | 0.05 |
| | ZeroR | 0.53 | 0.21 | 0.32 |
| | J48 | 0.91 | 0.84 | 0.07 |
| | RandomForest | 0.90 | 0.82 | 0.08 |

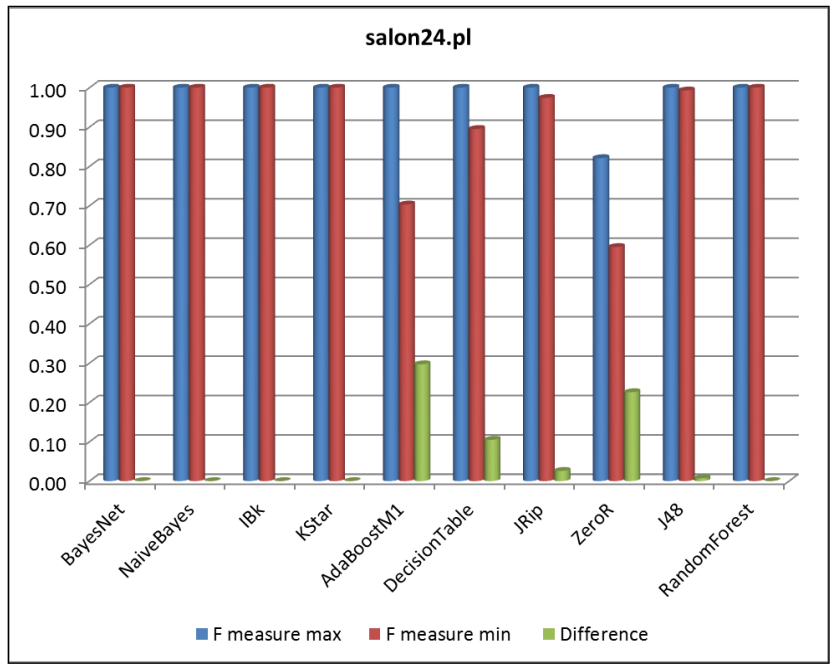

**Fig. 2.** The classifiers comparison for salon24.pl

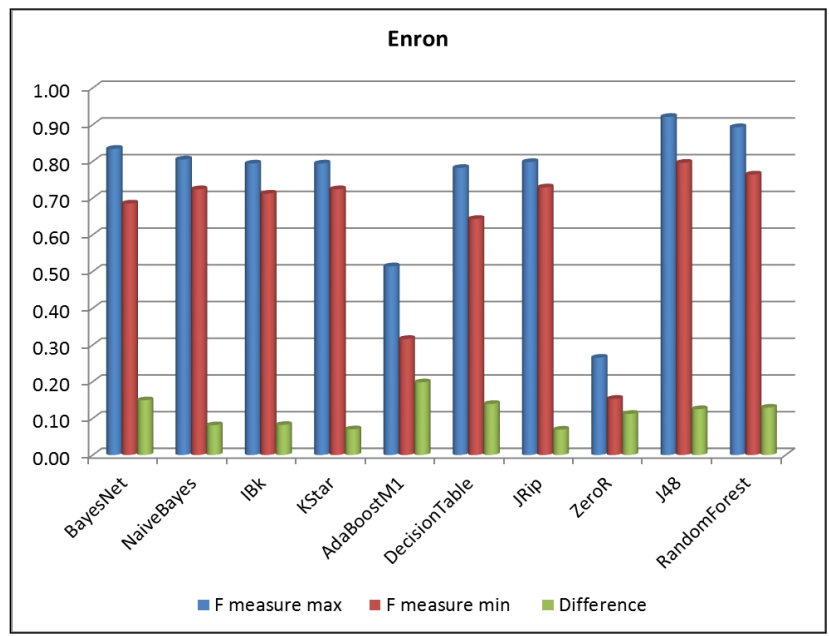

**Fig. 3.** The classifiers comparison for Enron

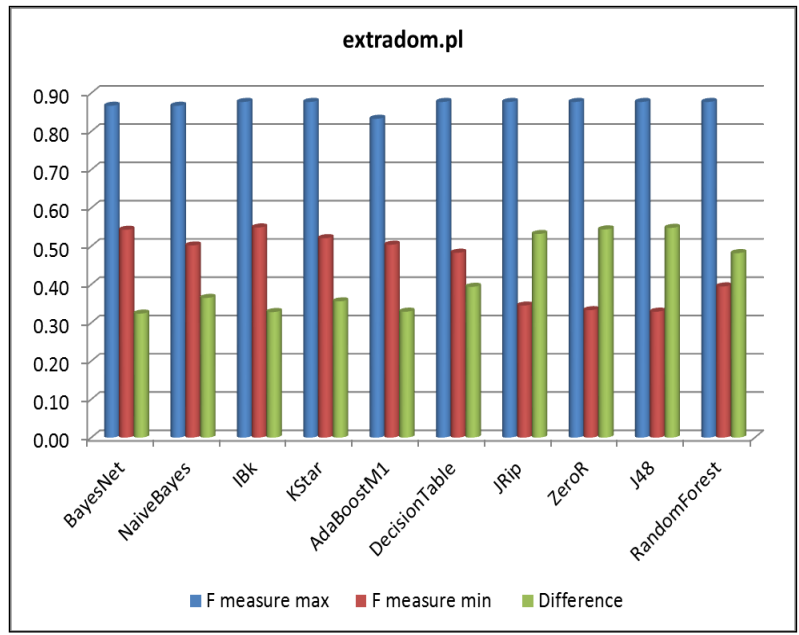

**Fig. 4.** The classifiers comparison for extradom.pl

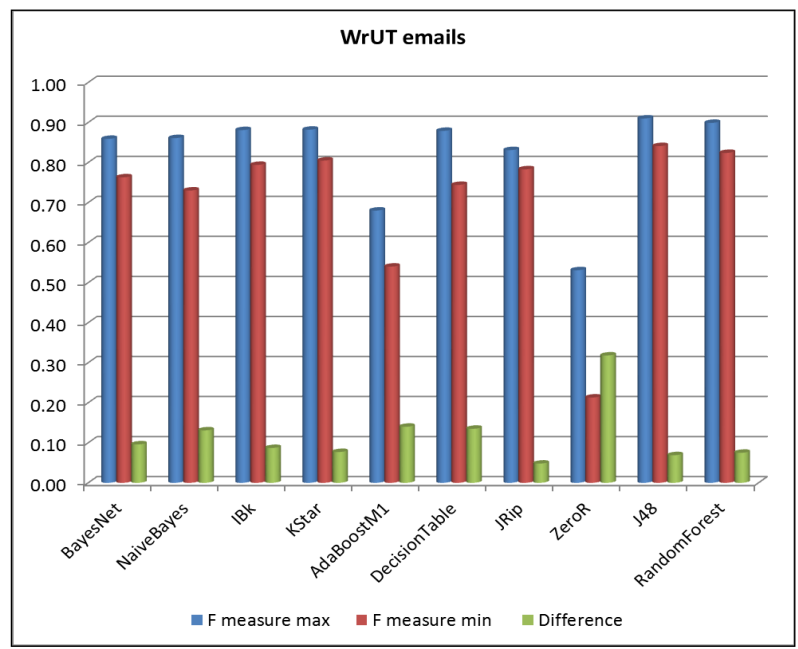

**Fig. 5.** The classifiers comparison for WrUT emails

Table 2 clearly indicates that for each dataset the best two classifiers are J48 (C4,5) decision trees and Random Forest ensemble of decision trees, thus, both classifiers were used for further analyses. Additionally, the results for these two classifiers are quite impressive since F measure for both of them is always around 0.8-0.9.

Now, it is necessary to analyse how the α and β parameters affect the classification. This was done for the WrUT dataset. The first analysis was for J48 and is presented in Table 3 and Figures 6, 7.

**Table 3.** The weighted average of F-measure measure (weighted by the contribution of the class–event in the dataset) for F48 decision tree for all six possible classes

| β\α [%] | 50 | 60 | 70 | 80 | 90 | 100 |
|---|---|---|---|---|---|---|
| 50 | 0.881 | 0.85 | 0.887 | 0.889 | 0.884 | 0.888 |
| 60 | 0.884 | 0.879 | 0.898 | 0.885 | 0.883 | 0.91 |
| 70 | 0.886 | 0.89 | 0.897 | 0.902 | 0.897 | 0.884 |
| 80 | 0.879 | 0.885 | 0.889 | 0.91 | 0.886 | 0.882 |
| 90 | 0.87 | 0.882 | 0.871 | 0.913 | 0.892 | 0.887 |
| 100 | 0.852 | 0.869 | 0.848 | 0.907 | 0.869 | 0.841 |

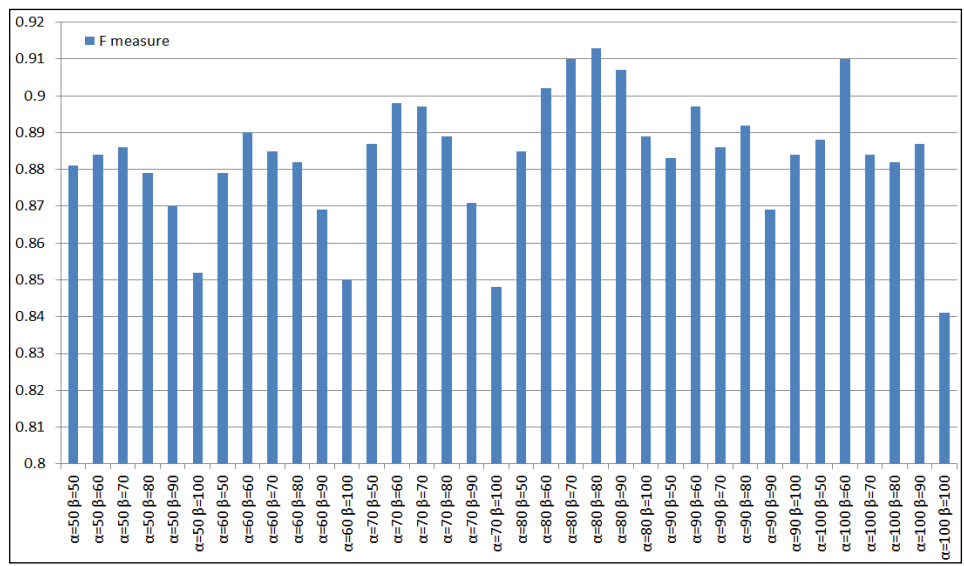

**Fig. 6.** F-measure values in relation to β and α

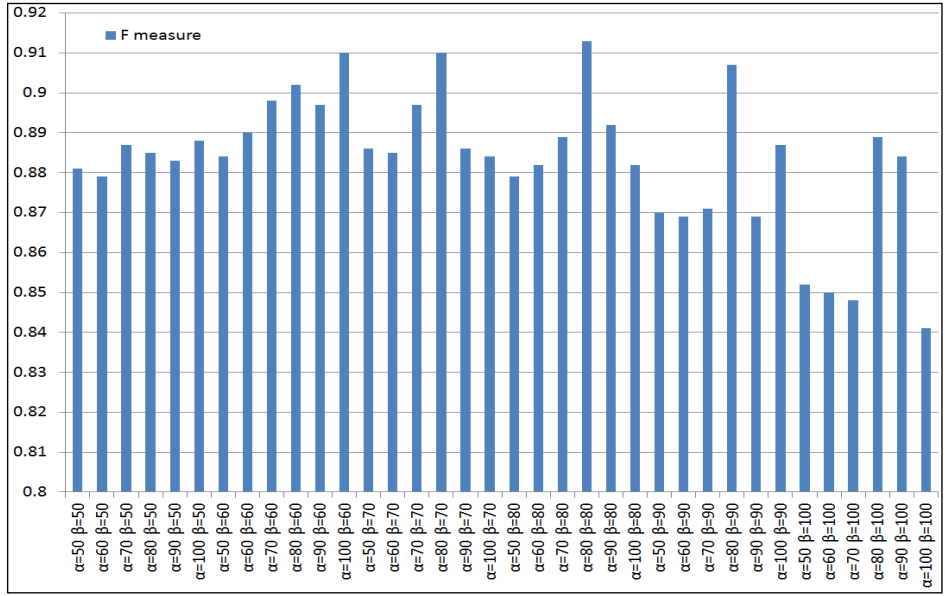

**Fig. 7.** F-measure values in relation to α and β

While analysing Figures 6 and 7 for the constant α, we can observe the best results are when β is around 80%. However, for the constant β, it is hard to see any regular pattern. In general, the highest F-measure is for α = 80%. So, if the J48 decision tree is used as a classifier, it is recommended to use α = 80% and β from the set {70%, 80%, 90%} for the *GED* method parameters. The reason behind such a result can be quite simple. If we look at results presented in [3] we can see that the high α and β reduce the number of split and merge events. Thus, the number of those events is similar to the number of other events. On the other hand, for the low α and β the number of splits and merges overshadow the number of the other events. It means that value of about 80% appears to be the best with respect to classification quality evaluated by the F-measure.

Quite similar results were achieved by the Random Forest classifier. The parameter α can be from the set {80%, 90%, 100%} and β from {60%, 70%, 80%, 90%}. Hence, the conclusion is: the *GED* method with the high α and β produces better input features for classification, also if applied to the Random Forest classifier. The evaluation of α and β influence with the Random Forest classifier was presented in Table 4, Figure 8 and 9. Not like for J48 tree, for Random Forest tree a specific pattern can be found for both α and β.For the constant α the best results are if β is equal to 60%, 70% or 80, see Figure 5.8, and for the constant β the best results are when α is equal to 80%, 90% or 100%, see Figure 9.

Table 4. The weighted average of F measure for Random Forest tree for all six classes

| β\α [%] | 50 | 60 | 70 | 80 | 90 | 100 |
|---|---|---|---|---|---|---|
| 50 | 0.846 | 0.848 | 0.857 | 0.874 | 0.868 | 0.87 |
| 60 | 0.848 | 0.852 | 0.865 | 0.881 | 0.875 | 0.899 |
| 70 | 0.846 | 0.853 | 0.872 | 0.891 | 0.879 | 0.897 |
| 80 | 0.849 | 0.854 | 0.862 | 0.893 | 0.882 | 0.867 |
| 90 | 0.843 | 0.848 | 0.849 | 0.896 | 0.872 | 0.887 |
| 100 | 0.828 | 0.824 | 0.828 | 0.869 | 0.869 | 0.849 |

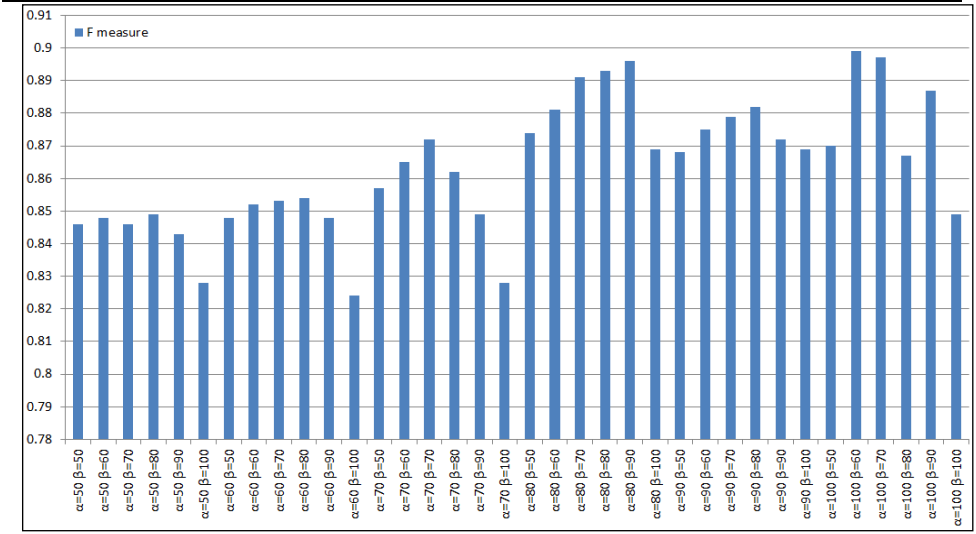

**Fig. 8.** F-measure values in relation to β and α

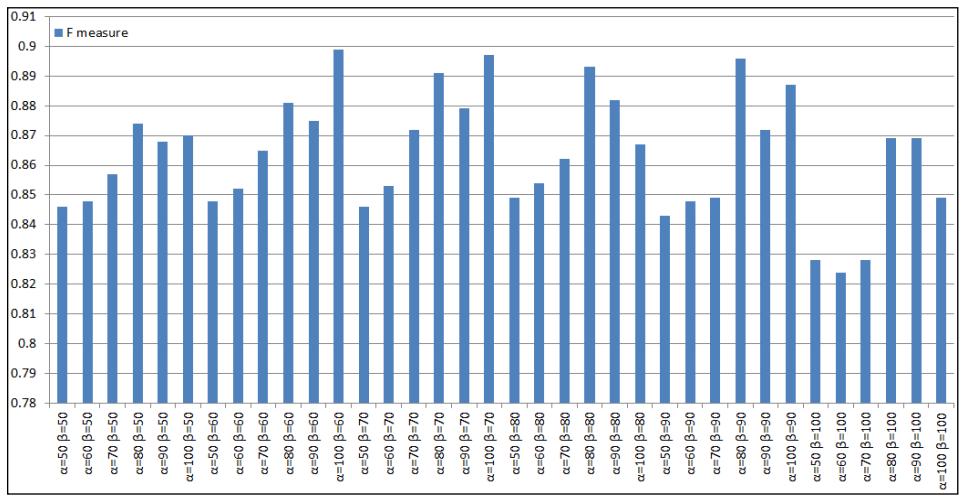

**Fig. 9.** F-measure values in relation to α and β

## 6   Conclusions and Future Work

It was shown that using a simple sequence which consists only of several preceding groups' sizes and events as an input for the classifier, the learnt model is able to produce very good results even for simple classifiers. It means that such prediction of group evolution can be very efficient in terms of prediction quality. The experimental analyses on six evolving social networks have revealed that decision tress and random forest as classifiers usually provide the most accurate results. Additionally, we can observe that the GED method used for change identification can be successfully used as a right indicator. However, its two parameters α and β significantly influence on the classification quality and the best results can be achieved for their values at the level of about 80%.

Of course, many questions remain unsolved, in particular:

- Are similar prediction results achievable for every kind of network?
- What would happen, if we use different classifiers or more advanced classification concepts like competence areas (clustering of groups and application of separate classifiers to each cluster)?
- What would be the influence of adding more input features (measures) describing the group like its diameter, average degree, percentage of network members which are in this group, the number of core members etc. as well as their various aggregations, e.g. average size for last 6 time frames?
- What would be the results, if we use shorter/longer sequences (more preceding events and group measures)?
- What would happen, if we use different node importance measure used in *GED*?

All above questions will be addressed in future research. This paper however, aimed only to present that predicting group evolution using the *GED* method with some common classifiers is both possible and effective.

### Acknowledgements

The work was partially supported by fellowship co-financed by the European Union within the European Social Fund and the Polish Ministry of Science and Higher Education, the research project 2010-13.